\documentclass[conference]{IEEEtran}
\IEEEoverridecommandlockouts
\usepackage{cite}
\usepackage{amsmath,amssymb,amsfonts}
\usepackage{algorithmic}
\usepackage{textcomp}
\usepackage{xcolor}
\usepackage[utf8]{inputenc}
\usepackage[pdftex]{graphicx}
\usepackage{subcaption}
\usepackage{comment}
\usepackage{float}
\usepackage{comment}
\usepackage{mathtools}
\usepackage[english]{babel}
\usepackage{amsthm}
\usepackage{amssymb,amsmath,amsthm}

\usepackage{graphicx}
\usepackage{svg}
\usepackage[pdftex]{graphicx}

\def\BibTeX{{\rm B\kern-.05em{\sc i\kern-.025em b}\kern-.08em
    T\kern-.1667em\lower.7ex\hbox{E}\kern-.125emX}}

\begin{document}

\title{Small-signal stability in inverter-dominated grids: exploring the role of gains, line dynamics, and operating conditions}
\author{
\IEEEauthorblockN{Ruth Kravis$^\star$, Gabriel E. Col\'on-Reyes$^\star$, Duncan S. Callaway$^{\star, \dagger}$}
\IEEEauthorblockA{$^\star$Department of Electrical Engineering and Computer Sciences, University of California, Berkeley, United States \\
$^\dagger$Energy and Resources Group, University of California, Berkeley, United States\\
\{ruth.kravis, gecolonr, dcal\}@berkeley.edu}}

\maketitle

\begin{abstract}

In a power grid with growing penetrations of renewable energy sources, inverters play a larger role in the dynamic interactions among network components. However, much is yet to be studied regarding inverter-dominated grid stability. This has initiated a re-examination of long-established assumptions made when conducting power systems studies. 

In this work we study the small-signal stability characteristics of a power system comprised of inverter and synchronous machine sources, and consider the effect of inverter control gains, transmission line dynamics, and network operating conditions.

We utilize \texttt{PowerSimulationsDynamics.jl} to perform this study on the IEEE WSCC 9 Bus test system. We vary inverter parameters across reported ranges in the literature, we consider three transmission line models of varying fidelity, and compare results across different operating conditions. 

We find that line dynamics are generally important for correctly assessing small-signal stability in networks with large penetrations of grid-following converters, and that gain tuning in these networks should utilize high fidelity line models.\\

\end{abstract}


\section{Introduction}
As we continue to integrate converter interfaced generation (CIG) into the grid, the potential for unstudied/under-studied dynamics to emerge increases. This is due to the influence of inverter control gains, and in particular, due to interactions of fast acting inverter control loops and fast electromagnetic dynamics of power systems \cite{hatziargyriou2020definition}. 

In this work, we study the small-signal stability of inverter-dominated grids, while varying three elements: (1) the operating condition, which we characterise as the combination of loading and share of load supplied by each generator \cite{markovic2021understanding}, (2) the gains/parameters of the generators, and (3) the transmission line model. For (1), we choose representative scenarios for an inverter-dominated grid, which includes a grid-forming (GFM), grid-following (GFL), and synchronous machine (SM). For (2), we focus on the gains of the GFM control loops, which include the virtual-synchronous machine (VSM) outer control loop, and the cascaded voltage and current inner control loops \cite{d2015virtual}. We use controller gains reported in the literature \cite{NREL_VSM_params}, and propose a method to bring gains from different sources together based on time-scale separation arguments. For (3), we consider three transmission line models of increasing fidelity: $statpi$: an algebraic $\pi$ line model, $dynpi$: a dynamic $\pi$ line model, and multi-segment single-branch ($MSSB$): a dynamic, segmented $\pi$ line model that captures higher frequency dynamics \cite{D’Arco_Beerten_Suul_2015_cable_MOR}. The higher fidelity  transmission line models will allow us to capture interactions between states that are not captured by lower fidelity models. 






We seek to answer the following research question:
 how do transmission line models affect small-signal stability conclusions under different operation conditions and gain choices? 

We find that: 1) line models matter most in operating conditions with high GFL share, and when they do matter, they have a stabilizing effect, 2) high GFL share conditions tend to be less stable, which is consistent with results reported in the literature, 3) under certain operating conditions, one set of gains will result in a mostly unstable system using one line model, but an always stable system using a different line model. This indicates that inverter gain tuning requires using the highest fidelity line model possible, particularly when working with GFL-dominated grids.  

\textit{Notation: }Dot notation indicates the time derivative of a variable, i.e., $\dot{x} = \frac{dx}{dt}$. Bold lower-case symbols are used to represent complex variables in the $dq$ reference frame, e.g. $\boldsymbol{x}=x_d + jx_q$.



\section{Models}

We use \texttt{PowerSimulationsDynamics.jl} (PSID) to model the system and perform small-signal stability analysis. 

\begin{figure*}[tb]
 \begin{subfigure}{.6\columnwidth}
    \includegraphics[width=.8\columnwidth]{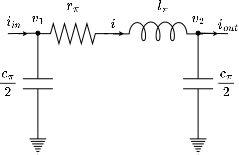}
    \caption{}
        \label{fig:pi_schematic}
  \end{subfigure}
    \begin{subfigure}{1.4\columnwidth}
        \centering
        \includegraphics[width=\columnwidth]{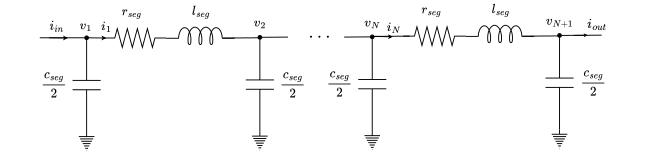}
        \caption{}
        \label{fig:ms_schematic}
    \end{subfigure}
    \caption{Line model schematics: (a) $\pi$ model ($statpi$ and $dynpi$), and (b) $MSSB$}
    \label{fig:line_models}
\end{figure*}

\subsection{Power System Components}
\subsubsection{Generator}
Generators are modeled with five main components: a stator, a shaft, a turbine governor, a power system stabilizer (PSS), and an automatic voltage regulator (AVR). The models we choose, as well as parameters, can be found in Chapters 15 and 16 of \cite{milano2010power}. We adopt the six-state Anderson-Fouad machine stator model. We choose a shaft model given by the swing equations with damping. We opt for a fixed input turbine governor, a Type 1 AVR, and we do not include a PSS. 
\subsubsection{Inverter}
Inverters are modeled with six main components: a DC voltage source, a model for the switches, an output filter, an outer GFM or GFL control loop, an inner control loop, and a frequency estimator. We choose models and parameters from \cite{Beerten_D’Arco_Suul_2016}. This includes a fixed DC voltage source model, an LCL filter, a VSM GFM model for the outer loop, nested proportional-integral (PI) loops for the inner control loops, and a phase locked loop (PLL) for damping of the virtual frequency. We choose an averaged model for the switches \cite{yazdani2010voltage}. For the GFL control mode, we use active and reactive power PI controllers for the outer loop, and a single inner current control PI loop \cite{kenyon2021open}. 
\subsection{Transmission lines}\label{sec:tx_models}
To model TLs of arbitrary length $\ell$, we use per unit length parameters for impedance, $z_{km} = r_{km} + j\omega l_{km}$, and shunt admittance, $y_{km} = g_{km} + j\omega c_{km}$. At a particular operating frequency $\omega$, we compute the lumped parameter equivalent $\pi$ model according to the following equations, which include hyperbolic correction factors from the steady state solution to Telegrapher's equations:
\begin{align}
    z_{\pi} &= z_{km} \ell \left( \frac{\sinh(\gamma \ell)}{\gamma \ell} \right) \label{eqn:hyperbolic1}\\ 
    y_{\pi} &= y_{km} \ell \left( \frac{\tanh(\gamma \ell/2)}{\gamma \ell/2} \right) \label{eqn:hyperbolic2}\\
    \gamma &= \sqrt{z_{km}y_{km}} \label{eqn:gamma}
\end{align}
where $\ell$ is line length (distinct from inductance $l$).
We assume $g_{km}=0$, which does not imply that $g_\pi = \mathrm{Re}(y_{\pi})=0$, but we choose to overwrite $g_\pi = 0$, since it is several orders of magnitude smaller than the other line parameters. From Equations~\eqref{eqn:hyperbolic1}--\eqref{eqn:gamma}, we can compute equivalent $r, l, c$ as $r_\pi=\mathrm{Re}(z_{\pi}), l_\pi= x_\pi/\omega = \mathrm{Im}(z_{\pi})/\omega, c_\pi= b_\pi/\omega = \mathrm{Im}(y_{\pi})/\omega$. 

\subsubsection{Algebraic $\pi$ model (statpi)}

The algebraic $\pi$ model has the form shown in Fig.~\ref{fig:pi_schematic}. It assumes that any voltage/current dynamics are stable and settle quickly, compared to other system dynamics. Therefore, the differential terms associated with the line capacitance and inductance are set to zero, giving:
\begin{align}
    \boldsymbol{i_{in}} &= \left(\frac{1}{z_\pi} + y_\pi\right)\boldsymbol{v_1} - \frac{1}{z_\pi}\boldsymbol{v_2} \\
    \boldsymbol{i_{out}} &= -\frac{1}{z_\pi}\boldsymbol{v_1} - \left(\frac{1}{z_\pi}+y_\pi\right)\boldsymbol{v_2} \label{eqn:statpi}
\end{align}
Since this model is purely algebraic, any dynamics on $\boldsymbol{i_1}$, $\boldsymbol{i_2}$, $\boldsymbol{v_1}$, and $\boldsymbol{v_2}$ that arise due to the interconnection of the line with dynamic devices will instantaneously appear at the other end of the line. 

\subsubsection{Dynamic $\pi$ model (dynpi)}
The dynamic $\pi$ model has the same structure as shown in Fig.~\ref{fig:pi_schematic}, but it includes dynamics on line current and bus voltage states:
\begin{align}
    \frac{l_{\pi}}{\omega_b
}\frac{d\boldsymbol{i}}{dt} &= (\boldsymbol{v_1} - \boldsymbol{v_2}) - z_\pi\boldsymbol{i} \\
    \frac{c_{\pi}}{2
} \frac{1}{\omega_b}\frac{d\boldsymbol{v_1}}{dt} &= (\boldsymbol{i_{in}} - \boldsymbol{i}) - y_\pi\boldsymbol{v_1} \\
    \frac{c_{\pi}}{2}\frac{1}{\omega_b}\frac{d\boldsymbol{v_2}}{dt} &= (\boldsymbol{i} - \boldsymbol{i_{out}}) - y_\pi \boldsymbol{v_2} \label{eqn:dynpi} 
\end{align}

Similar to $statpi$, $dynpi$ only captures the effect of the distributed line parameters in steady state.

\subsubsection{Multi-segment single-branch $\pi$ model ($MSSB$)}

We define a line with multiple `segments' as one divided into a discrete set of $N$ identical length $\pi$ components in series. The parameters for each segment are given by:
\begin{align}
    r_{seg} &= r_{km}\ell_{seg} \\
    l_{seg} &= l_{km}\ell_{seg} \\
    c_{seg} &= c_{km}\ell_{seg} 
\end{align}
where $\ell_{seg} = \frac{\ell}{N}$. Further, $z_{seg} = r_{seg} + j\omega l_{seg}$ and $
y_{seg} = j\omega c_{seg}$. This model is shown in Fig. \ref{fig:ms_schematic}. The $i^{th}$ segment of an $N$-segment $MSSB$ model is defined by the following equations: 
\begin{align}
    \frac{l_{seg}}{\omega_b
}\frac{d\boldsymbol{i_{i}}}{dt} &= (\boldsymbol{v_i} - \boldsymbol{v_{i+1}}) - z_{seg}\boldsymbol{i_{i}} \\
    \frac{c_{seg}}{2}\frac{1}{\omega_b}\frac{d\boldsymbol{v_i}}{dt} &= (\boldsymbol{i_{i-1}} - \boldsymbol{i_{i}}) - y_{seg}\boldsymbol{v_i} \\
    \frac{c_{seg}}{2}\frac{1}{\omega_b}\frac{d\boldsymbol{v_{i+1}}}{dt} &= (\boldsymbol{i_{i}} - \boldsymbol{i_{i+1}}) - y_{seg}\boldsymbol{v_{i+1}} \label{eqn:mssb} 
\end{align}

Note that for $i=1$, $i_{i-1}=i_{in}$, and for $i=N$, $i_{i+1}=i_{out}$. As $N$ is increased, the $MSSB$ model more closely approximates the equivalent $\pi$ model in steady state frequency response. The advantage of explicitly representing segments is that unlike the equivalent $\pi$ model (Eqn.~\eqref{eqn:hyperbolic1}--\eqref{eqn:gamma}), it captures the distributed nature of the line parameters in both transient and steady state responses. 

\subsection{Line parameters}
To obtain line parameters, we start with the frequency dependent line impedance parameters from Table 3 in~\cite{Dommel_1985}. From this data, we build a frequency dependent model of $z_{km}$ via vector fitting~\cite{gustavsen1999rational, Beerten_D’Arco_Suul_2016}, a form of parameter estimation that incorporates real transmission line data. We then evaluate this model at nominal frequency to obtain $z_{km}$. In combination with a constant value for $y_{km}$, we use this value in Eqns.~\eqref{eqn:hyperbolic1}--\eqref{eqn:gamma} to obtain the lumped parameter model parameters for $statpi$ and $dynpi$. This approach is useful because it allows frequency dependent models to be built if needed, such as those in \cite{Beerten_D’Arco_Suul_2016} and \cite{Colon-Reyes_Kravis_Sharma_Callaway_2023}.

\subsection{Aggregate system model}

All the models described above can be written as a system of differential-algebraic equations (DAEs). Linking the devices and network, we arrive at a mathematical form as follows:
\begin{equation}
    \begin{bmatrix}
    \dot{x} \\ 0
\end{bmatrix}
= 
\begin{bmatrix}
    f(x, y, u) \\ g(x, y, u)
\end{bmatrix}
\end{equation}
Here, $x \in \mathbb{R} ^n$ are the system's dynamic states, $y \in \mathbb{R} ^m$ are the system's algebraic states, $u \in \mathbb{R} ^p$ are the system inputs. $f:\mathbb{R}^n \times \mathbb{R}^m \times \mathbb{R}^p \rightarrow \mathbb{R}^n$ and $g:\mathbb{R}^n \times \mathbb{R}^m \times \mathbb{R}^p \rightarrow \mathbb{R}^m$ are the vector equations associated to the dynamics of the network and the algebraic constraints. 

We are interested in studying the small-signal stability of such a system. In general, $f$ is a nonlinear vector field. Therefore, to study small-signal stability we find an equilibrium point $(x^\star, y^\star, 0)$ by setting $\dot{x} = 0$ and solving the nonlinear system of equations, then linearize around that point, and arrive at a set of linear dynamics that characterize the behavior of the system in the vicinity of that equilibrium point. The resulting equations will then be of the following form:
\begin{equation}
    \Delta\dot{x} = J(x^\star, y^\star) \Delta x
\end{equation} 

Here, $J(x^\star, y^\star)$ is the reduced system Jacobian matrix. By studying the eigenvalues of this matrix, we can determine if $(x^\star, y^\star, 0)$ is a stable or unstable operating condition for the network. See \cite{henriquez2020grid} or \cite{lara2023powersimulationsdynamicsjl} for further details on the linearization process in PSID.

\section{Small-Signal Stability Analysis}

\subsection{Test case}

\begin{figure}[]
    \centering
    \includegraphics[scale=0.37]{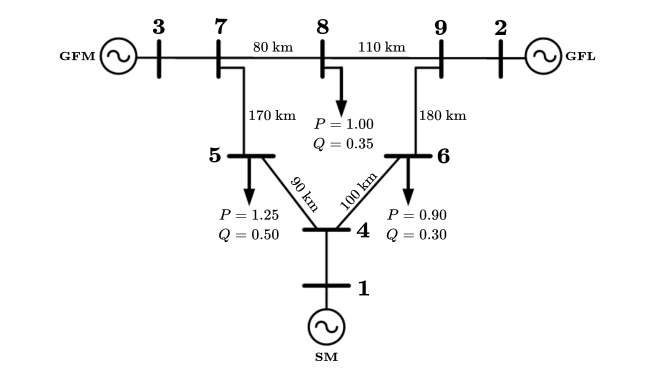}
    \caption{Modified IEEE 9 bus test system. SM at bus 1 (reference bus), GFL at bus 2, VSM GFM at bus 3.}
    \label{fig:9bus_schematic}
\end{figure}

We use the IEEE WSCC 9 Bus Test case, shown in Fig.~\ref{fig:9bus_schematic} to perform our analysis. We verify that the operating conditions we study in the same network, where each generator is a SM.

\subsubsection{Parameters}\label{sec:params}

We fix the parameters for the GFL, with outer loop PI gains shown in Table \ref{tab:GFL_params}. We use the  parameters from \cite{d2015virtual} for the inner loop, filter, and PLL. 

\begin{table}[h]
    \centering
    \begin{tabular}{ccc}
        \hline 
        \textbf{Parameter} & \textbf{Symbol} & \textbf{Value [p.u.]}  \\ \hline \hline 
        Active power control P gain & $k_{pp}$ & 0.0059   \\
        Active power control I gain & $k_{ip}$& 7.36 \\
        Reactive power control P gain & $k_{pq}$& 0.0059  \\
        Reactive power control I gain & $k_{iq} $& 7.36  \\ 
        \hline 
    \end{tabular}
    \caption{GFL parameters}
    \label{tab:GFL_params}
\end{table}

Table~\ref{tab:params} summarises the GFM gain parameters we chose to vary. Ranges are constructed from values reported in  \cite{d2015virtual}, \cite{markovic2021understanding}, and \cite{NREL_VSM_params}, although we note that a much wider range of parameters is reported across the literature. 

\begin{table}[h]
    \centering
    \begin{tabular}{ccccc}
        \hline 
        \textbf{Parameter} & \textbf{Symbol} & \textbf{Range} & \textbf{Unit} \\ \hline \hline 
        VSM inertia time constant & $T_a$ & 0.5 -- 2 & s \\
        VSM damping coefficient & $k_d$ & 100 -- 400 & p.u. \\
        Reactive Power Droop gain &  $k_q$ & 0.05 -- 0.2 &  p.u.\\
        Inner voltage loop P gain &  $k_{p,v}$ & 0.5 -- 0.6 &  p.u.\\
        Inner voltage loop I gain & $k_{i,v}$ & 400 -- 800 & p.u.\\
        Inner current loop P gain & $k_{p,c}$ & 0.74 -- 1.27 & p.u.\\
        Inner current loop I gain & $k_{i,c}$ & 1.19 -- 14.3 & p.u.\\
        \hline 
    \end{tabular}
    \caption{GFM parameter ranges}
    \label{tab:params}
\end{table}

\begin{figure*}[tb]
\centering
    \begin{subfigure}[t]{0.24\textwidth}
    \includegraphics[width=\textwidth]{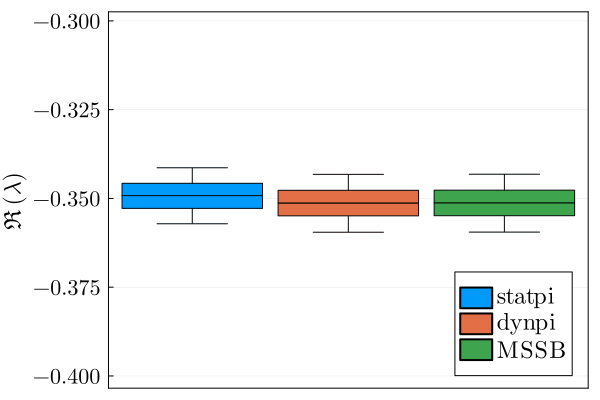}
    \caption{}
        \label{fig:low_load_case1}
    \end{subfigure}
    \begin{subfigure}[t]{0.24\textwidth}
        \centering
        \includegraphics[width=\textwidth]{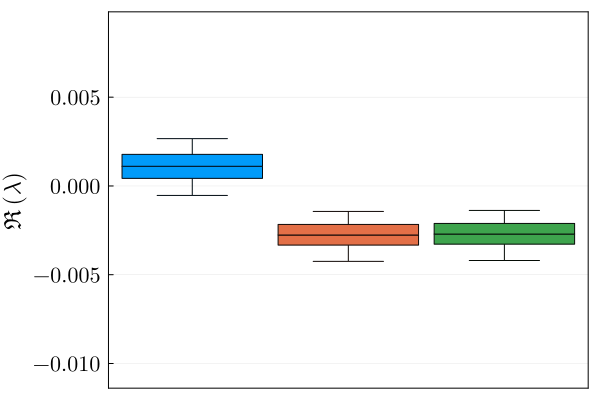}
        \caption{}
        \label{fig:low_load_case2}
    \end{subfigure}
    \begin{subfigure}[t]{0.24\textwidth}
        \centering
        \includegraphics[width=\textwidth]{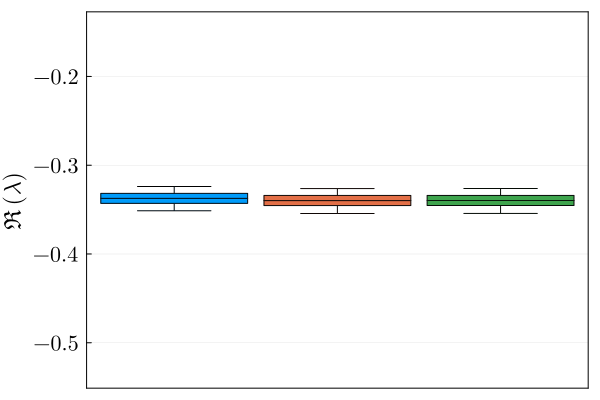}
        \caption{}
        \label{fig:low_load_case3}
    \end{subfigure}
    \begin{subfigure}[t]{0.24\textwidth}
    \includegraphics[width=\textwidth]{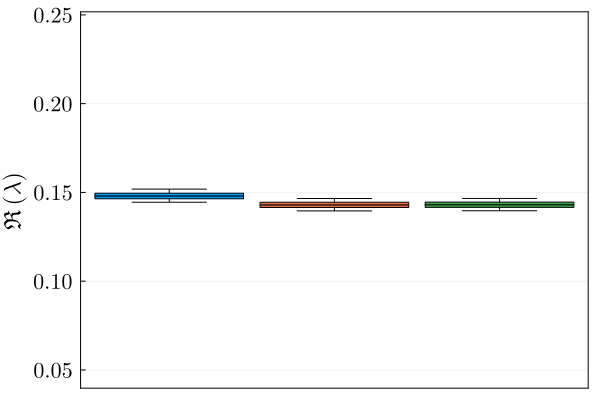}
        \caption{}
        \label{fig:low_load_case4}
    \end{subfigure}
    \vfill
        \begin{subfigure}[t]{0.24\textwidth}
    \includegraphics[width=\textwidth]{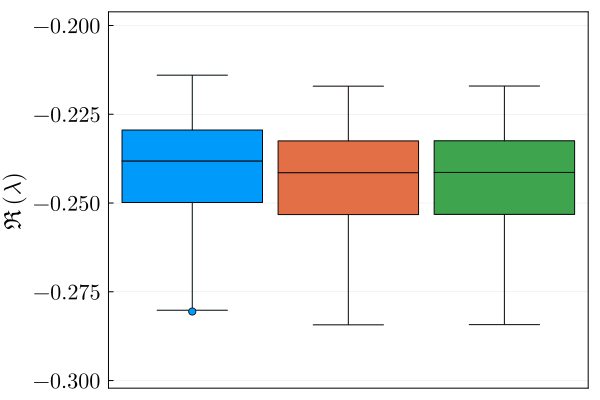}
    \caption{}
        \label{fig:nom_load_case1}
    \end{subfigure}
    \begin{subfigure}[t]{0.24\textwidth}
        \centering
        \includegraphics[width=\textwidth]{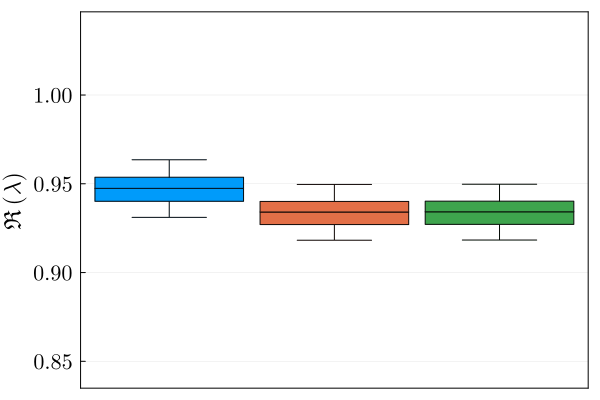}
        \caption{}
        \label{fig:nom_load_case2}
    \end{subfigure}
    \begin{subfigure}[t]{0.24\textwidth}
        \centering
        \includegraphics[width=\textwidth]{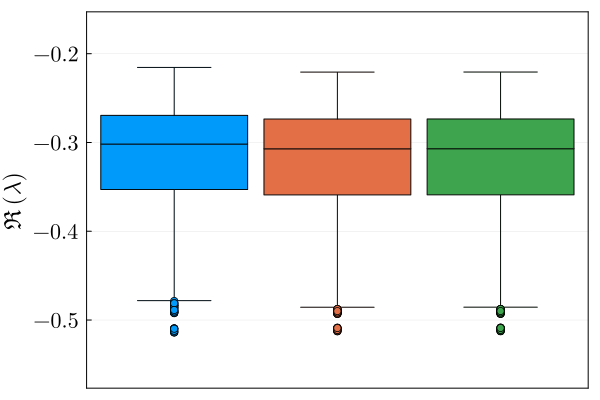}
        \caption{}
        \label{fig:nom_load_case3}
    \end{subfigure}
    \begin{subfigure}[t]{0.24\textwidth}
    \includegraphics[width=\textwidth]{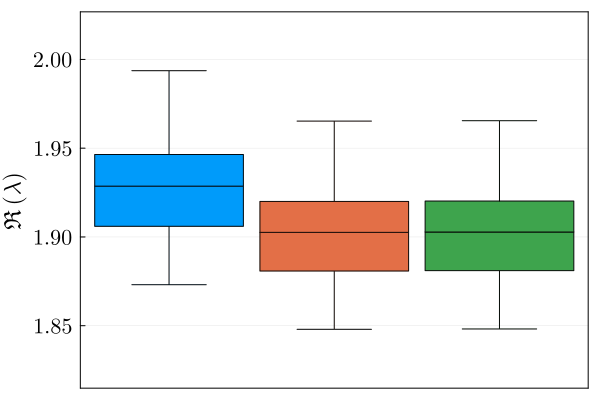}
        \caption{}
        \label{fig:nom_load_case4}
    \end{subfigure}
    \caption{Boxplots for different operating conditions, showing the real part of the least stable eigenvalue for different line models. \textbf{Top row} is $load \ scale=0.4$, \textbf{bottom row} is $load \ scale = 1.0$. \textbf{Columns} correspond to Case 1 through 4, respectively. All columns except for (b)/(f) have the same y-axis range (difference between maximum and minimum value of the eigenvalue).} 
    \label{fig:2x4_cases}
\end{figure*}
Not all combinations of parameters are desirable from an engineering perspective. For example, separation of time scales is an important attribute of GFM and GFL control architecture design. Therefore, we filter the parameters based on engineering considerations.  

For the GFM controls, we require that that $k_{pc}$ is $10 \times$ larger than $k_{pv}$. The same requirement is applied to the GFL: $k_{pc}$ must be $10 \times$ larger than $k_{pp}$ and $k_{pq}$. This will help achieve time scale separation. We also require that the corresponding integral gains for all these controllers be $10 \times$ larger than the proportional gain to help with driving steady state errors to zero.

\begin{table}[h]
    \centering
    \begin{tabular}{cccc}
    \hline 
        \textbf{Case Name} & $\mathbf{\eta_{ibr}}$ & $\mathbf{\eta_{gfl}}$ & $\mathbf{\eta_{gfm}}$  \\ \hline \hline 
        Case 1 & 70\% & 14\% & 56\% \\ 
        Case 2 & 70\% & 42\% & 28\%  \\ 
        Case 3 & 90\% & 18\% & 72\% \\ 
        Case 4 & 90\% & 54\% & 36\% \\ 
         \hline 
    \end{tabular}
    \caption{Operating conditions}
    \label{tab:operating_conditions}
\end{table}  

Lastly, we construct operating scenarios where inverters supply a high (70\%) and very high (90\%) share of total active power. We also vary which inverter is injecting more active power. Table~\ref{tab:operating_conditions} summarises the operating conditions analysed -- IBR refers to inverter-based resources. The values quoted in the table refer to the fraction of active power load supplied by the GFM and GFL. Losses are picked up by the SM at the slack bus, so in reality, these fractions are slightly smaller than what is quoted. Note that $\eta_{ibr} = \eta_{gfm} + \eta_{gfl}$.
 
Finally, we choose two $load \ scale$ values, to simulate light and heavier loading. We conduct experiments by first discretizing the gain choices according to our filtering scheme, then randomly sampling 1000 parameter choices from a possible 7000 combinations. Then, for each set of sampled parameter choices, and for each case and $load \ scale$, we perform small-signal stability analysis of the system using all three line models. We examine the real part of the least stable eigenvalue to evaluate stability at that operating point.


\subsection{Results and Discussion}

\begin{figure*}[]
 \begin{subfigure}[t]{0.33\textwidth}
    \includegraphics[width=\textwidth]{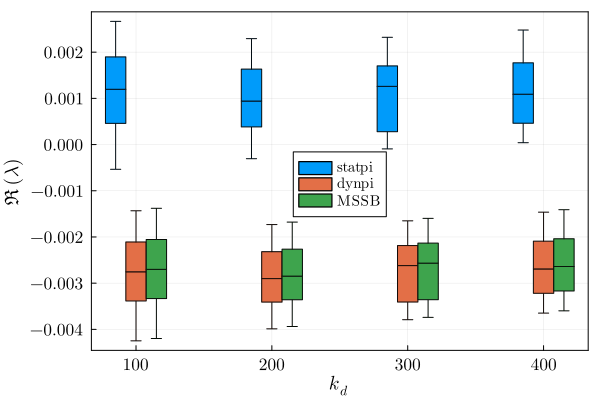}
    \caption{$k_d$}
        \label{fig:kd_case}
  \end{subfigure}
    \begin{subfigure}[t]{0.33\textwidth}
        \centering
        \includegraphics[width=\textwidth]{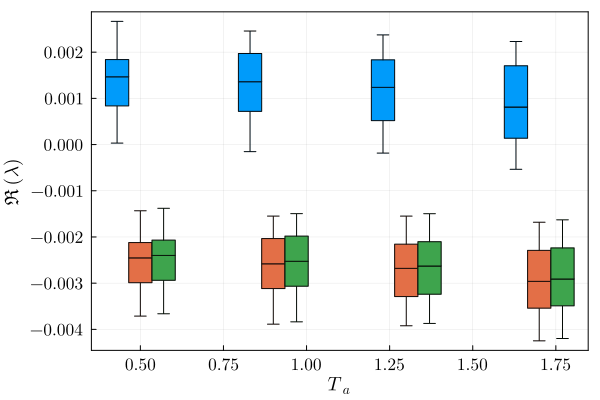}
        \caption{$T_a$}
        \label{fig:Ta_case}
    \end{subfigure}
    \begin{subfigure}[t]{0.33\textwidth}
        \centering
        \includegraphics[width=\textwidth]{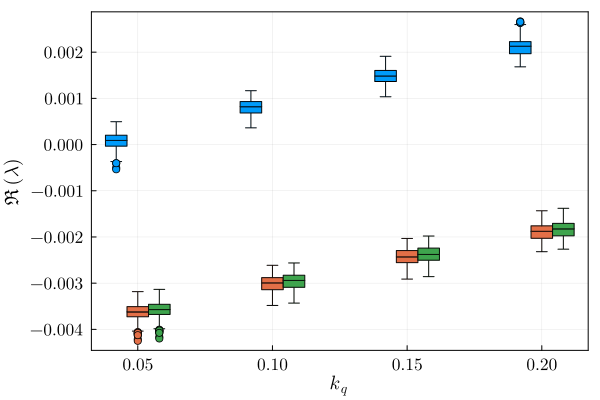}
        \caption{$k_{q}$}
        \label{fig:kq_case}
    \end{subfigure}
    \caption{Boxplots for outer loop GFM gains, with $load\ scale=0.4$, case 2 (corresponding to Fig.~\ref{fig:low_load_case2}).}
    \label{fig:gain_boxplots}
\end{figure*}

\subsubsection{Effect of generation share}
Fig.~\ref{fig:2x4_cases} presents results for the eight operating conditions selected. Firstly, we observe that high GFL share cases (column 2 and 4) are less stable for a fixed $\eta_{ibr}$ (i.e., comparing column 1 to 2, and 3 to 4). This effect is more pronounced for larger $\eta_{ibr}$. These results agree with our expectations and prior results \cite{markovic2021understanding}.


\subsubsection{Effect of loading}
We observe that the range in the eigenvalues' real part is much larger for heavier loading. Furthermore, for some gain choices, increasing the loading actually results in a more stable system. For example, comparing Fig.~\ref{fig:low_load_case3} and ~\ref{fig:nom_load_case3} we see that Fig.~\ref{fig:nom_load_case3} (higher loading) can be either more or less stable than Fig.~\ref{fig:low_load_case3} (lower loading). This suggests that -- unsurprisingly -- under high loading gains must be chosen carefully.

\subsubsection{Effect of line models}
Including line dynamics has either no effect or a stabilizing effect on the equilibrium point, and this effect is most pronounced for high $\eta_{gfl}$. This means $statpi$ provides a conservative estimate of system stability. This is seen clearly in Fig.~\ref{fig:low_load_case2}, where using $statpi$ may result in an incorrect assessment of stability. In other cases (e.g., Fig.~\ref{fig:nom_load_case2}) line dynamics do not always provide sufficient stabilizing to move the eigenvalues to the LHP. Echoing our conclusions on generation share, since $statpi$ and $dynpi$/$MSSB$ are most different at high $\eta_{gfl}$, and $MSSB$ is the highest fidelity line model, this suggests that $statpi$ should be treated with caution at high GFL shares.  

\subsubsection{Effect of gains}
We explore Fig.~\ref{fig:low_load_case2} to understand why $statpi$ and $dynpi$/$MSSB$ are different under high GFL shares, and if there are any trends in the gains that might give insight. Fig.~\ref{fig:gain_boxplots} shows boxplots for the three outer loop GFM gains that were varied in this analysis ($k_d$, $T_a$, and $k_q$). The inner loop gains ($k_{pv}$, $k_{iv}$, $k_{pc}$, and $k_{ic}$) were also swept, but they didn't have a strong trend/influence on the overall stability of the system so are not included here. 

In Fig.~\ref{fig:kq_case} we see a strong upward trend, which suggests that, for the ranges of parameters sampled, $k_q$ has the most influence on small-signal stability. This result could be a consequence of the parameter ranges we chose -- other gains may exhibit trends over a larger range of values. 

\begin{figure}
    \centering
    \includegraphics[width=0.45\textwidth]{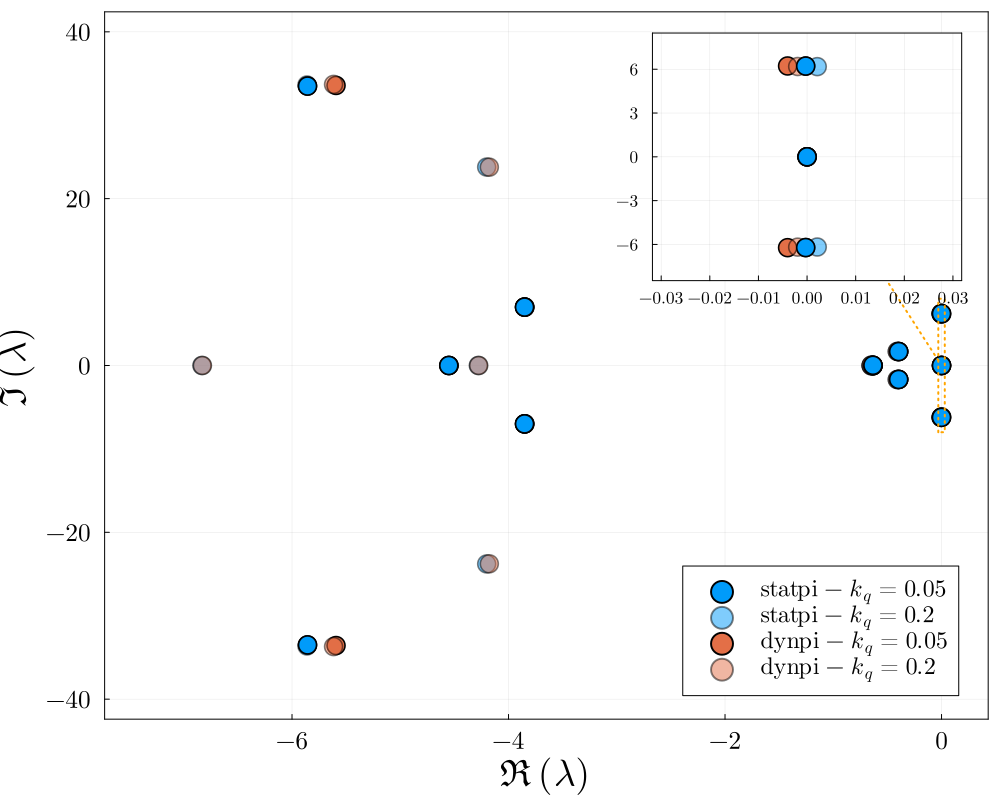}
    \caption{Eigenvalues for $load\ scale=0.4$, case 2 (Fig.~\ref{fig:low_load_case2}), for two different sets of parameters. Solid points correspond to $k_q=0.05$ (both line models stable), light points correspond to $k_q=0.2$ ($statpi$ unstable, $dynpi$ stable).}
    \label{fig:eig_plot}
\end{figure}

Lastly, Fig.~\ref{fig:eig_plot} examines how the eigenvalues in Fig.~\ref{fig:low_load_case2} change for two different parameter sets, one where $k_q=0.05$ and one where $k_q=0.2$. For the set where $k_q=0.05$, the system is stable for $statpi$ and $dynpi$ (solid points). For the set where $k_q=0.2$, the system is unstable for $statpi$ but stable for $dynpi$. The eigenvalue pair that cross into the RHP are strongly participated in by the the GFL $\sigma_q$ state, as measured by participation factors. This state  corresponds to the GFL outer loop reactive power control. The unstable eigenvalues are also complex, which has implications for appropriate measures for estimating stability (e.g., a damping ratio may be appropriate). 

\section{Conclusions and Future Work}

We find that transmission line model choice can alter a modeler's conclusion about a system's small-signal stability in inverter-dominated grids.  This is especially true for GFL-dominated systems.  In the cases we studied we found that lower fidelity line models (those without dynamics) tended to be more conservative with respect to small-signal stability. This suggests that using low fidelity line models will not lead to false positive characterizations of system stability. In this work we also see that the operating condition significantly affects the overall stability of the system, regardless of line models. We also see that the parameter choice for GFM gains matters more under heavier loading, in particular, the reactive power outer loop gain $k_q$. We also observed that unstable eigenvalues are complex, motivating further exploration of appropriate measures for estimating stability in inverter-dominated grids.


Future work includes integrating GFL gains into the parameter ranges. We also saw a large range of reported parameters for inverters, including parameters that do not adhere to the engineering principles discussed in Section~\ref{sec:params}. This large range, as well as the diversity of test systems studied in the literature, makes comparing results across studies challenging. Further work should focus on establishing sets of parameters and test systems that can enable better benchmarking. 




\section*{Acknowledgment}
This research was supported by the U.S. Department of Energy's Solar Energy Technologies Office through award 38637 (UNIFI Consortium).
\bibliographystyle{IEEEtran}
\bibliography{references}

\end{document}